# Toward Compact Fiber In-line Nonlinear Devices via Highly Efficient Nanophotonic Cavity Interface

*Mahsa Haddadi Moghaddam*[1,†], *Kirlie Iulius Figuera Michal*[1,†], *Sangwoo Lee*[1], *Sijin Sung*[1], *Jongwon Lee*[2], *Hyeong-Ryeol Park*[1], *and Je-Hyung Kim*[1,3]*

[1]Department of Physics, Ulsan National Institute of Science and Technology, Ulsan 44919, Republic of Korea

[2]Department of Electrical Engineering, Ulsan National Institute of Science and Technology, Ulsan 44919, Republic of Korea

[3]Graduate School of Quantum Science and Technology, Ulsan National Institute of Science and Technology, Ulsan 44919, Republic of Korea

Corresponding author: Je-Hyung Kim jehyungkim@unist.ac.kr

†Mahsa Haddadi Moghaddam and Kirlie Iulius Figuera Michal contributed equally to this work.

**Abstract**

Compact and efficient frequency conversion within optical fibers is highly desirable for nonlinear and quantum photonic technologies, yet it remains challenging due to weak nonlinear interactions and limited coupling efficiencies onto optical fibers. Here, we demonstrate resonantly enhanced second-harmonic generation (SHG) through the all-fiber integration of a gallium nitride (GaN) hole-type circular Bragg grating (h-CBG) cavity, directly transferred onto a standard optical fiber. Together with the large second-order nonlinear susceptibility and wide optical transparency window of GaN, the fabricated h-CBG membrane cavity on GaN enables strong field confinement and vertically directional out-coupling of the generated SHG signal. As a result, we observe drastically enhanced SHG signals from the h-CBG device compared with the bulk GaN and the unpatterned freestanding GaN membrane. Using a deterministic pick-and-place transfer technique, we demonstrate robust and precise fiber integration of the GaN cavity device, enabling in-line SHG generation from a conventional fiber platform. This work establishes a compact and scalable approach for incorporating optical nonlinearity into fiber-based photonic systems.

## 1. Introduction

Nonlinear optical processes provide routes for frequency conversion, signal processing, and the generation of nonclassical optical states, such as entangled or squeezed states, making them essential for both classical and quantum technologies.[1-6] As photonics systems advance toward scalable and deployable architectures, there is a growing demand to implement these nonlinear processes directly within practical photonics platforms, particularly optical fibers. Fiber-based implementation offers several advantages of eliminating bulky free-space alignment and ensuring stable and scalable signal delivery over a long distance with minimized loss. Among high-order nonlinear processes, second-harmonic generation is one of the most fundamental and widely utilized frequency-conversion processes, enabled by the second-order nonlinear susceptibility ($\chi^2$) of non-centrosymmetric materials.[7-10] Bulk nonlinear crystals, such as *β*-barium borate, have been numerously explored with their high $\chi^2$ values.[11-12] However, a strict phase-matching condition is required to sustain coherent nonlinear interaction over long propagation lengths. To relax these constraints and improve conversion efficiency, periodically poled techniques have been widely developed.[13-14] Despite their performance, these bulk platforms remain difficult to integrate with optical fibers because their millimeter-scale dimensions lead to significant mode mismatch and coupling loss, and their strong in-plane anisotropy requires precise polarization control. Recently, ultrathin two-dimensional nonlinear materials have emerged as promising alternatives for compact and integrable nonlinear optics.[15-17] Their atomic-thin thickness relaxes conventional phase-matching requirements and enables broadband operation, making them attractive for direct integration with optical fibers.[18-22] However, the inherently limited interaction volume in such ultrathin media and the small light collection angle of optical fibers significantly restrict nonlinear conversion efficiency and fiber coupling efficiency.

To address these limitations, introducing engineered photonic structures, such as metasurfaces [23-24], nanowires [25-26], and cavities [27-30] integrated with strong nonlinear materials, provides a promising route toward compact and fiber-integrable nonlinear optical devices. Especially, photonic cavities can effectively increase the local optical field, and their subwavelength thickness relaxes conventional phase-matching requirements for broadband operation.[31] Furthermore, photonic structures tailor the spatial mode profile to match the guided mode of an optical fiber and add the controlability over the phase, amplitude, and polarization. Among a variety of thin-membrane type cavities, an h-CBG cavity is especially well suited for fiber integration,[32] and they have demonstrated fiber-based single-photon arrays [33] and fiber-quantum sensors. [34] The radially engineered hole arrays support strong field confinement at the cavity center together with highly directional far-field characteristics. However, realizing efficient cavity integration on thin nonlinear materials together with practical fiber integration remains challenging.

In this work, we demonstrate compact and efficient in-line generation of a second-harmonic signal in a conventional fiber platform using an integrated gallium nitride (GaN) membrane cavity. GaN features a wide bandgap,[35-37] a large optical window from UV to NIR,[38-39] a large $\chi^2$ nonlinear coefficient,[40-41] and excellent robustness under high optical power.[42-43] In addition, GaN can be grown as a wavelength-thick thin membrane, which is highly advantageous for planar nanofabrication, precise cavity patterning, and transfer-based integration onto diverse photonic platforms, such as chip and fiber systems. Taking advantage of these material platforms, we fabricate an h-CBG cavity on a GaN membrane on a silicon substrate and achieve highly efficient SHG with a normalized conversion efficiency of $3.62\times10^{-3}$ $W^{-1}$, corresponding to a near 32-fold enhancement compared with an unpatterned freestanding thin GaN membrane. Using a deterministic pick-and-place transfer technique, we directly integrate the h-CBG cavity onto the fiber facet with high alignment to the fiber core, enabling efficient mode coupling and stable fiber in-line SHG operation. Therefore, our approach provides a scalable and practical route toward fiber-integrated nonlinear photonic devices.

## 2. Results and discussion

**Figure 1a** shows a schematic of the fiber in-line nonlinear device, where a GaN cavity is placed between two optical fibers for in-line optical pumping and signal collection of the generated SHG. Unlike previous fiber-integrated nonlinear materials without optimal photonic interfaces, our approach realizes an efficient fiber in-line SHG configuration by integrating a GaN h-CBG membrane cavity directly within optical fibers. In this architecture, the h-CBG plays an important dual role as a resonant nonlinear cavity for SHG enhancement and as a directional light-guiding structure for efficient coupling into a low-numerical-aperture fiber, thereby overcoming the major limitations of earlier approaches and providing enhanced conversion efficiency, relaxed phase-matching requirements, and stable, controllable in-line operation.

### 2.1. Cavity device simulation and fabrication

To enhance efficiencies for both generation and collection of SHG, we optimized the cavity design through numerical simulations based on the finite-difference time-domain (FDTD) method. The simulated structural parameters included the radial period between holes (r), the radius of the central disk (h), the hole size (a), and the axial period between holes (b) (see the schematic in Supporting Information S1). The h-CBG, rather than conventional ring-type CBGs, features a highly directional out-coupling and a high $Q$ value of a cavity mode. In addition, its interconnected membrane structure, compared with conventional ring-type CBGs, enables robust and reliable cavity transfer while minimizing geometric deformation and structural damage during the device transfer process.[32-34] The cavity mode was optimized for 784 nm pump wavelength and supports high directionality at both the pumping and SHG wavelengths, enabling efficient excitation and collection through the fiber.

Figure 1b presents the simulated far-field radiation pattern of the cavity mode as a function of the divergence angle at the pump wavelength, while Figure 1c illustrates the corresponding directional emission derived from the simulated three-dimensional far-field radiation pattern shown in Figure 1b. Figure 1d shows the FDTD-simulated near-field electric field intensity ($|E|^2$) distribution overlaid on the SEM image of the fabricated GaN h-CBG cavity, demonstrating the strong optical confinement within the cavity. These simulated results show that the h-CBG provides strong radial confinement, predominantly concentrated in the cavity center, and produces a vertically directed beam.

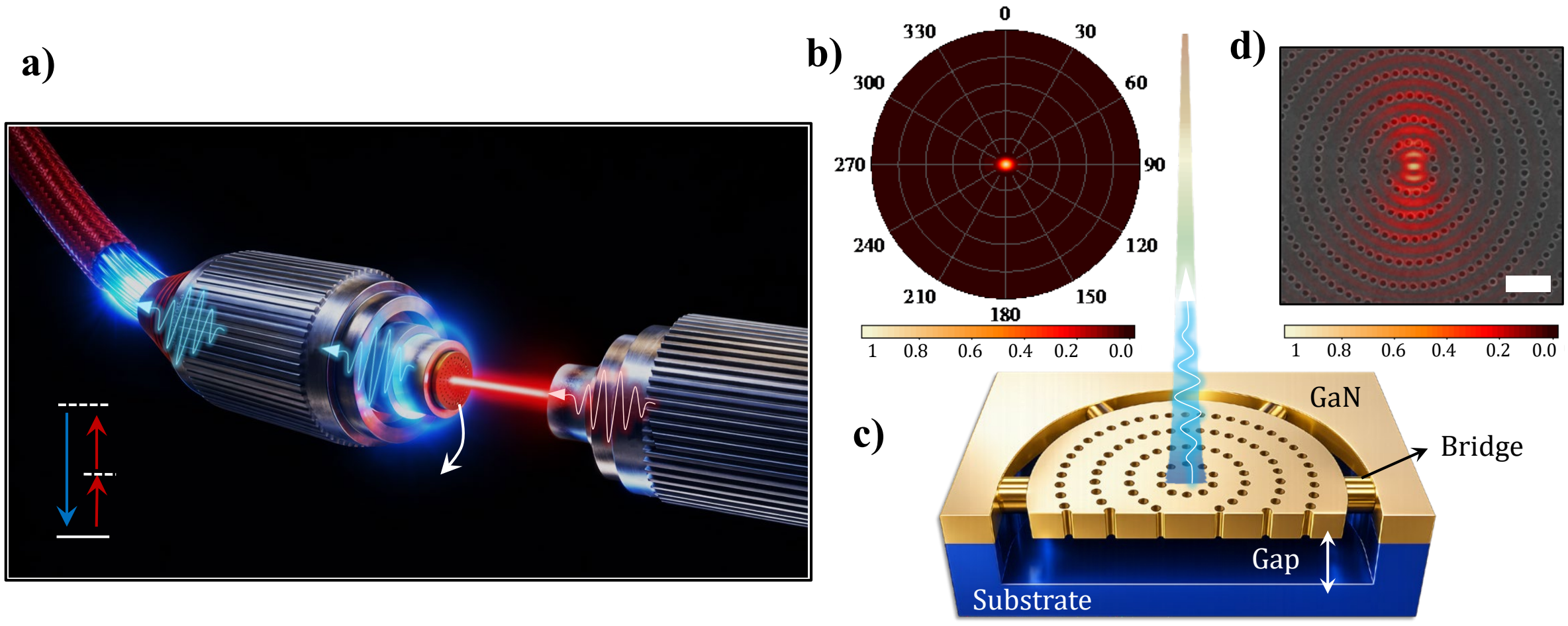

**Figure 1.** (a) Schematic of the fiber-integrated frequency-conversion device, in which a GaN h-CBG is positioned on the fiber facet to enable excitation and collection of the second-harmonic signal. (b) Simulated far-field radiation profile of the cavity mode. (c) Schematic illustration of the cavity mode's directional emission derived from the simulated three-dimensional (3D) far-field radiation pattern shown in (b). (d) SEM image of the fabricated GaN h-CBG cavity overlaid with the simulated near-field cavity mode profile (scale bar is 1 μm).

From the simulated cavity designs, we fabricated the cavity on a 350 nm-thick GaN thin film on a Si substrate using electron beam lithography, followed by a dry etching process. To create an air-suspended membrane, we conducted a wet etching process for the silicon substrate. Surrounding micrometer-scale bridge structures temporarily connect the cavity to the parent substrate and are designed to fracture during device release and transfer onto the fiber core. Further fabrication details are provided in the Supporting Information S2.

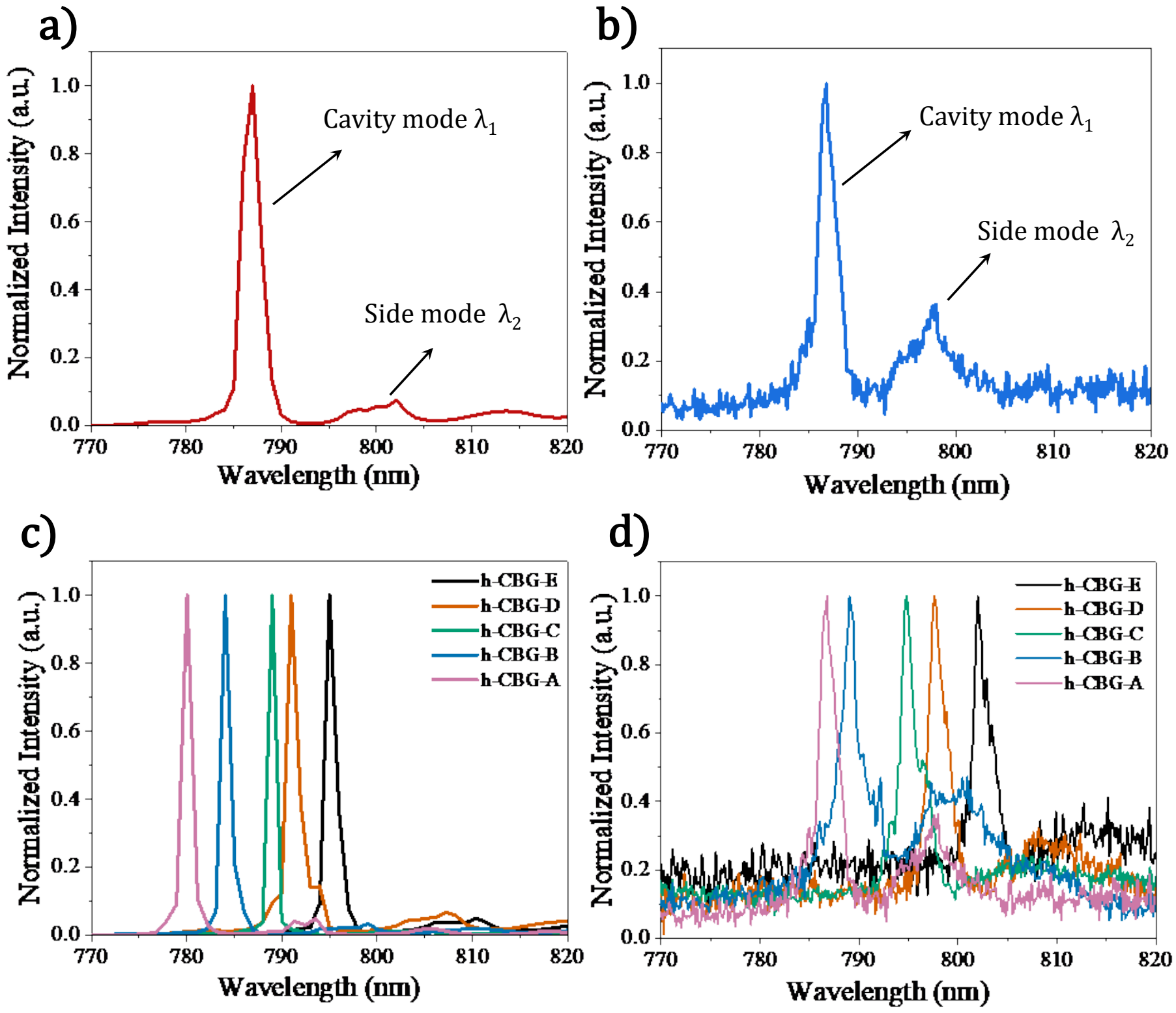

**Figure 2.** (a) Simulated and (b) measured cavity-mode intensity of h-CBG-B as a function of wavelength. (c) Simulated cavity-mode spectra for h-CBGs with different geometric parameters listed in Table S2, and (d) the corresponding experimental mode spectra.

**Figure 2a** and 2b compare the simulated and experimentally measured cavity mode spectra. In the experiment, the cavity spectrum of the fabricated GaN h-CBG cavity was characterized using a custom-built cross-polarized white-light spectroscopy setup at room temperature (see Supporting Information S3), where the cavity signal is isolated from the directly reflected background through cross-polarization detection. Both the simulation and measured spectrum exhibit a strong and narrow resonance at 784 nm, corresponding to the target pump wavelength. The simulated and measured resonance linewidths show a quality factor $(Q \approx \frac{\lambda}{\Delta\lambda})$ of 364 and 345, respectively. Therefore, the cavity mode of the fabricated device is well-matched to the simulation result. Together with bright cavity modes, a small side peak appears in both simulated and measured spectra of the h-CBG. This peak corresponds to a higher-order mode, but does not have a Gaussian and vertically directional field profile. An important aspect of a photonic cavity is the ability to tune the resonance wavelengths through geometric design. The resonant frequency of the cavity mode can be controlled by adjusting cavity designs, such as hole size and their radial and lateral periods. To scan the cavity mode across the 780~800 nm range, we independently optimized the cavity parameters at each wavelength. The detailed optimized cavity geometries were described in Table S4 of the Supporting Information. Figure 2c and 2d compare simulated cavity-mode spectra with different geometric parameters, and the corresponding measured mode spectra. The measured cavity spectrum shows the expected cavity mode shift, and a slight enlargement of the fabricated structure relative to the nominal design leads to a corresponding redshift in the resonance wavelength.

### 2.2. SHG performance in GaN h-CBG

We next confirm the generation of SHG from the GaN cavity device. **Figure 3a** schematically illustrates the enhanced SHG observed from the resonant GaN h-CBG device at the cavity wavelength along with the corresponding SHG signal at the doubled frequency. In the experiment, we pumped the GaN cavity using a femtosecond pulsed laser centered at 783.6 nm with a full width at half maximum (FWHM) of 7.01 nm, using a home-built optical measurement system described in Supporting Information Section S5. The generated SHG signal from the GaN cavity device was collected using an objective lens and directed to a spectrometer through a dichroic mirror and a band-pass filter to prevent scattered pump light from entering the spectrometer and producing second-order diffraction artifacts. Figure 3b shows representative measured spectra of the pump laser and the SHG signal, together with the corresponding Gaussian fits shown as red and blue lines, respectively. A pronounced SHG peak is observed at 391.76 nm, with a linewidth nearly half that of the pump, providing clear evidence of efficient SHG through frequency doubling. Group-III nitride films grown on silicon crystallize in the wurtzite phase, for which the nonzero second-order nonlinear tensor components are limited by symmetry to $\chi^{(2)}_{zzz}$, $\chi^{(2)}_{zxx}$, and $\chi^{(2)}_{zyy}$, with the usual symmetry between the last two indices. In this work, the observed second harmonic generation is primarily governed by the $\chi^{(2)}_{zzz}$ component, corresponding to the $d_{33}$ coefficient, which is aligned with the cavity mode resonant at the pump wavelength.[31, 44] Under this condition, efficient SHG occurs when both energy and angular momentum are conserved, giving rise to the expected relation $\lambda_{SHG} = \frac{\lambda_{Pump}}{2}$. As shown in the log-log plot in Figure 3c, the SHG intensity exhibits a clear quadratic dependence on the pump power over the range of 10 ~ 300 mW for an incident pulse width of 130 fs. The fitted slope of 1.99 ± 0.02 is in close agreement with the theoretical value, confirming the second-order nonlinear nature of the process. Figure 3d compares the collected SHG spectra measured from the GaN thin film on the substrate, the freestanding GaN film, and the GaN h-CBG cavity. Although GaN possesses a relatively large second-order nonlinear susceptibility, second-harmonic generation in a GaN film on silicon remains extremely weak due to limited effective interaction length.[36-37, 45] After undercutting the silicon layer to form an air-suspended GaN membrane, we measured an observable SHG spectrum. This enhancement arises from improved optical confinement within the air-clad GaN membrane, together with reduced absorption and improved extraction of the SHG signals. Despite this improvement, the nonlinear interaction strength and signal collection in the freestanding membrane remain small. In contrast, by introducing the h-CBG cavity onto the membrane, the local pump field intensity is dramatically enhanced within a small mode volume of the cavity, and multiple resonance in the coupled cavity mode enable coherent buildup of the generated SHG field within the thin nonlinear medium. Moreover, the h-CBG cavity supports highly directional vertical emission, which is advantageous for collecting the SHG signals. As a result, the GaN h-CBG cavity exhibits a significant SHG enhancement of approximately 32 times compared with the freestanding GaN membrane. Under an average pump power of 93 mW, the GaN h-CBG cavity produces a power-normalized efficiency of $3.62\times10^{-3}$ $W^{-1}$ (see Supporting Information S6 for further details). This result clearly shows the critical role of an optical cavity for efficient nonlinear processes in thin membrane structures.

The long-term stability of the SHG output was evaluated under continuous high-power irradiation at a pump power of 93 mW for one hour (Supporting Information S7). The SHG intensity remained highly stable throughout the measurement, with negligible variation from its initial value (see the inset of Figure S7 for the temporal evolution of the SHG intensity). This result confirms the excellent repeatability and operational robustness of the cavity-enhanced SHG process, highlighting the potential of semiconductor nanocavities as stable and reliable nonlinear light sources.

Figure 3e shows the polarization-dependent SHG response measured at the resonant pump wavelength for both the freestanding GaN film and the GaN h-CBG nanocavity. In this measurement, the y-polarized SHG component was monitored while the polarization of the excitation laser rotated from 0° (y-axis) to 360°. For the freestanding GaN membrane, the SHG intensity exhibits a four-fold rotational symmetry, consistent with the crystalline symmetry of wurtzite GaN.[38, 46] In contrast, the SHG response from the GaN h-CBG nanocavity displays a pronounced two-fold polarization dependence.[23, 31] This change indicates that the polarization characteristics are strongly influenced by the cavity modes beyond the intrinsic nonlinear response of GaN. Although the h-CBG geometry is rotationally symmetric, slight elongation or fabrication-induced asymmetry in the patterned hole array can introduce significant optical anisotropy, leading to polarization-selective resonant enhancement. Consequently, the result implies that the polarization properties of the generated SHG can be engineered through careful cavity design.

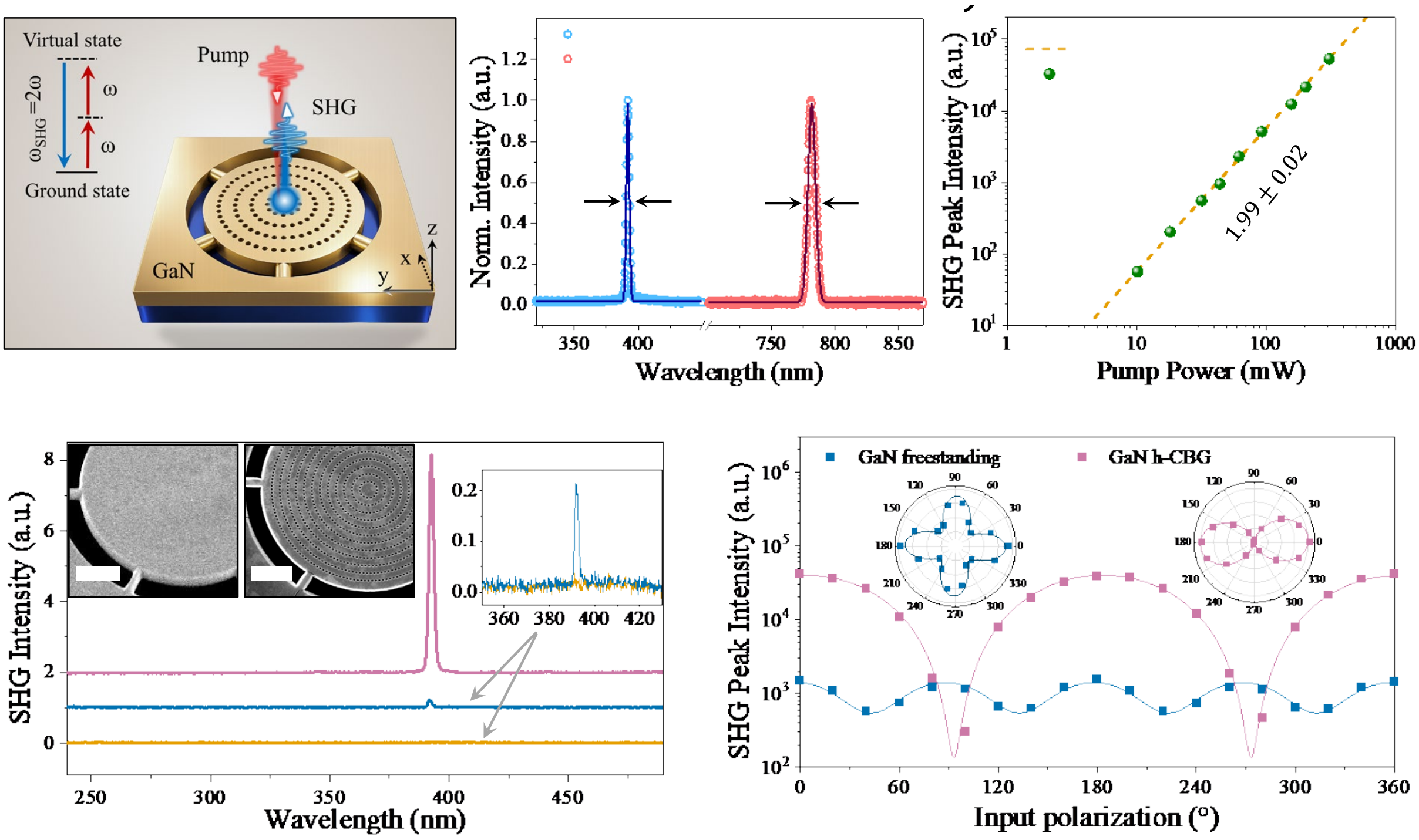

**Figure 3.** (a) Schematic illustration of the SHG generation from a GaN h-CBG membrane cavity. (b) Normalized spectra of the fundamental pump laser (red) and the generated SHG signal (blue) from the GaN h-CBG cavity. (Solid lines represent Gaussian-fitted curves) (c) Measured SHG intensity as a function of pump power, showing the power dependence of the nonlinear response. (d) Comparison of SHG signals from (I) the GaN h-CBG cavity, (II) the freestanding GaN membrane, and (III) the GaN on the silicon substrate. All scale bars are 2 μm. **(e)** Polarization-dependent SHG intensity measured on and off the cavity as a function of pump polarization angle, where 0° corresponds to polarization parallel to the y-axis. The insets show polar plots of the corresponding polarization-dependent SHG intensity.

### 2.3. Resonant Enhancement of SHG in the GaN h-CBG Cavity

As the cavity strongly governs the SHG response of the GaN membrane, resonantly pumping the cavity device is an important condition. To achieve precise resonance matching for enhanced SHG, we examined the SHG response with continuous spectral evolution of the cavity mode over the 780~800 nm range with a fixed pumping wavelength. The cavity mode (top) and corresponding SHG (bottom) spectra from three different spectral detunings between GaN h-CBG and the pumping laser are presented in **Figure 4a**, 4b, and 4c. For ease of comparison, all spectra are plotted symmetrically with respect to the pump and SHG wavelengths, and the pump spectrum is indicated by a yellow dashed line. The measured SHG intensity exhibits a strong dependence on the spectral detuning between the pump wavelength, $\lambda_p$, and the fundamental cavity resonance, $\lambda_1$, defined as $\Delta\lambda = |\lambda_p - \lambda_1|$. This behavior represents that the field enhancement occurs when the cavity mode spectrally overlaps with the pumping laser. Figure 4d plots the SHG intensity as a function of $\Delta\lambda$. When the cavity mode is nearly resonant with the pump wavelength ($\Delta\lambda \approx 0.56$), corresponding to the blue-shaded region, the SHG intensity reaches its maximum, as shown in Figure 4a (top and bottom). When the fundamental cavity mode is slightly detuned by less than approximately 6 nm ($\Delta\lambda \approx 3.63$ nm), partial spectral overlap with the pump is maintained. In this green-shaded region, a substantial SHG signal is still observed, as illustrated in Figure 4b, with a power-normalized conversion efficiency of $1.09\times10^{-5}$ $W^{-1}$. By contrast, for cavities with a larger detuning of more than 6 nm, the fundamental cavity mode no longer overlaps spectrally with the pump. In some cases ($\Delta\lambda \approx$ 9.12 nm), SHG can still be generated through coupling to the cavity side mode, $\lambda_2$, as indicated by the pink-shaded region in Figure 4d. However, because this side mode provides weaker field enhancement compared to the fundamental resonance, the resulting SHG signal is substantially reduced.

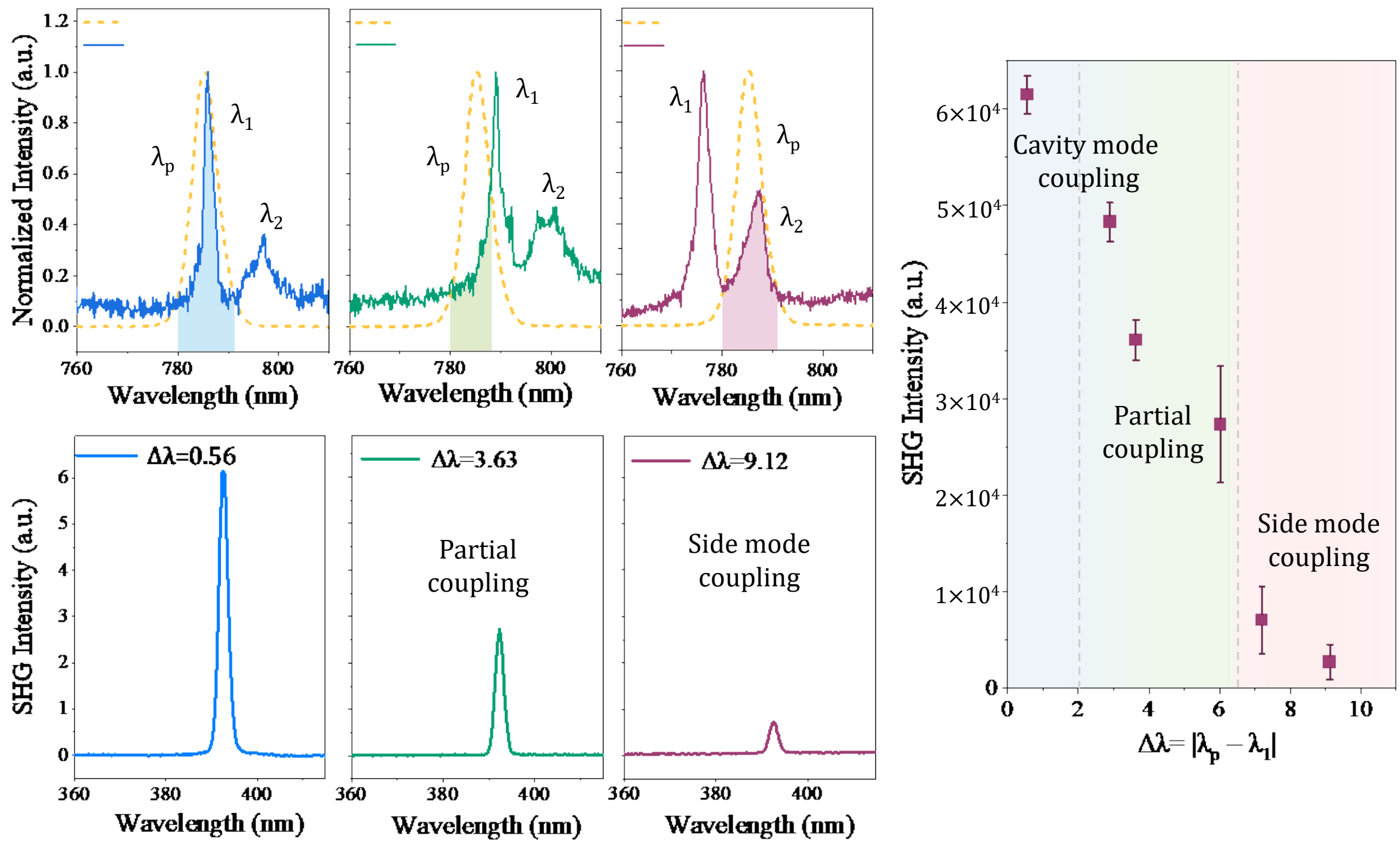

**Figure 4.** Cavity-mode spectra (top) and corresponding SHG intensity (bottom) for different spectral detuning ($\Delta\lambda = |\lambda_p - \lambda_1|$) between the cavity resonance and the pump wavelength: (a) Δλ=0.56 nm, (b) Δλ=3.63 nm, and (c) Δλ=9.12 nm. The yellow dashed line represents the pump spectrum, and the shaded regions indicate the spectral overlap between the cavity resonance and the fundamental pump mode. (d) Comparison of the SHG intensity as a function of spectral detuning.

### 2.4. SHG from in-line fiber integrated GaN h-CBG cavity

Finally, we demonstrate SHG from the fiber with an integrated GaN h-CBG cavity. The h-CBG membrane cavity has two important roles for fiber integration and interfacing. First, the membrane cavity provides a transferable nonlinear photonic device that is directly compatible with the end facet of a conventional optical fiber. **Figure 5a** illustrates the pick-and-place process using a micro polydimethylsiloxane (PDMS) stamp. During the picking process, the surrounding bridge-supported suspension structures are intentionally broken, allowing the cavity to separate from the original substrate and to be precisely positioned on the fiber core under an optical microscope (See supporting information S2). The pre-characterized GaN h-CBG device exhibiting the strongest SHG signal was selected and deterministically integrated onto a standard single-mode optical fiber. A step-by-step description of the integration method, with corresponding optical microscope images, is provided in Supporting Information S8.

Another important role of the h-CBG cavity is its capability to couple the generated second harmonic signals into the single-mode fiber. Together with strong field confinement in a cavity, the cavity supports a highly directional vertical far-field profile, allowing efficient coupling of both the pump and generated SHG signals directly into the fiber without additional optical components. Therefore, the integrated cavity-fiber architecture simultaneously enables efficient nonlinear generation and efficient signal delivery within a fiber platform.

To demonstrate the SHG from a fiber-integrated GaN cavity device, we first characterized the precisely integrated h-CBG cavity on the core of a single-mode fiber (SM300 fiber) while we sent and focused the pump laser on to fiber-integrated h-CBG cavity in free space using an objective lens. As shown in Figure 5b, a strong and well-defined SHG peak with an excellent Gaussian profile was obtained, confirming efficient second-harmonic generation from the fiber-integrated device. The SHG spectrum obtained from the fiber-integrated device closely matches that of the original on-chip h-CBG cavity, confirming that the transfer process preserves the cavity's nonlinear optical performance.

More importantly, we then carry out both excitation and collection of the SHG signal from an all-fiber-integrated GaN cavity configuration using an in-line fiber setup as illustrated in Figure 5c. An ultra-high-NA fiber (UHNA3) was employed for optical pumping, while the generated SHG signal was collected through the same SM300 fiber. As shown in Figure 5(d), a clear SHG peak is observed from the in-line fiber-integrated GaN h-CBG cavity, whereas no measurable SHG signal was detected from the bare fiber used as the reference. The fiber-integrated GaN h-CBG device achieved a power-normalized conversion efficiency of $5.32 \times 10^{-4}\ \mathrm{W}^{-1}$. Compared with the free-space excitation and fiber-collection configuration, the all-fiber in-line device exhibits a lower SHG efficiency. We attribute this reduction to the lower pumping power density delivered through the fiber compared with the tightly focused excitation by an objective lens. Nevertheless, the successful demonstration of efficient in-line fiber SHG generation establishes the nanophotonic cavity as a highly efficient optical interface that simultaneously enhances nonlinear frequency conversion and couples the generated signals directly into optical fibers. In this fiber in-line device, the pumping and collection fibers can be readily connected using standard mating sleeves, providing a compact, scalable, and plug-and-play frequency-conversion module.

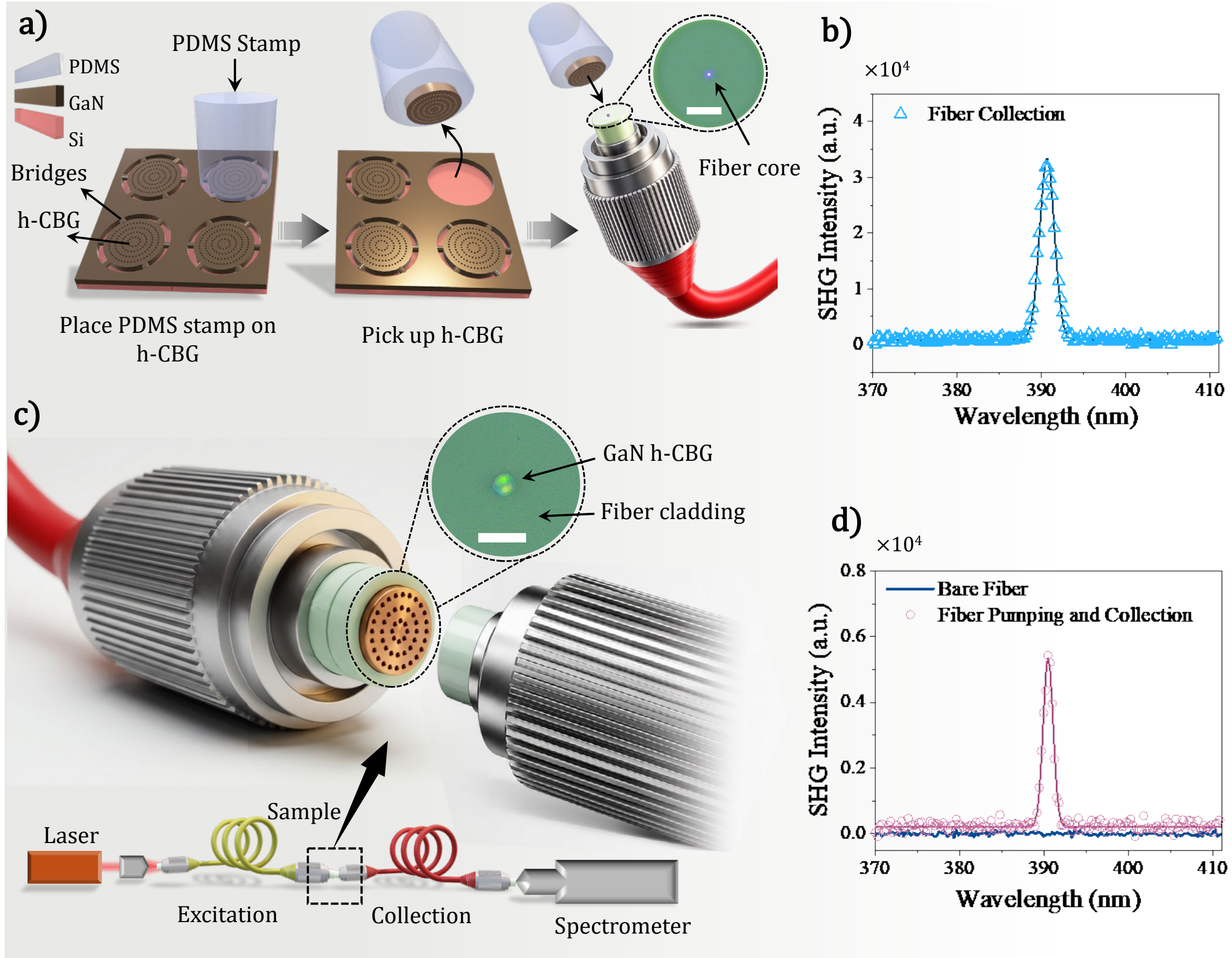

**Figure 5. (a)** Schematic illustration of the pick-and-place process used to transfer the h-CBG cavity onto the facet of an optical fiber. The inset shows an optical microscope image of the fiber core illuminated by a 403 nm laser coupled from the opposite end of the fiber. **(b)** SHG spectrum measured under free-space excitation, where the pump laser was focused onto the h-CBG cavity at the fiber core using an objective lens. The dark blue line represents the Gaussian-fitted curves. **(c)** Schematic of the fiber in-line measurement setup, showing the fiber-to-fiber contact configuration at the sample position. The inset presents an optical microscope image of the h-CBG cavity integrated onto the fiber core. **(d)** SHG spectrum measured under all-fiber excitation, where the pump laser was launched through the fiber to excite the h-CBG cavity and the generated SHG signal was collected through the fiber. The dark pink line represents the Gaussian-fitted curves. (All scale bars in the optical microscope images of the fiber facets are 20 μm.)

## 3. Conclusion

In conclusion, we have demonstrated a compact, efficient in-line fiber frequency conversion platform by deterministically integrating a GaN h-CBG cavity onto the facet of a standard single-mode optical fiber. This strategy overcomes a key limitation of existing fiber-integrated nonlinear devices by establishing efficient coupling between the fundamental and second-harmonic waves, enabling continuous enhancement of SHG along the transmission path. Unlike direct frequency-conversion within a fiber that relies on long interaction lengths or specialized fiber structures [ref?], our approach efficiently mates a cavity-integrated nonlinear device that does not require long fiber lengths or any fiber modification, but it offers improved usability together with significantly enhanced nonlinear interaction. The resulting device achieves a conversion efficiency of $5.32 \times 10^{-4}$ $W^{-1}$ and efficient coupling into low-numerical-aperture single-mode fibers. Our work paves a new way for a fiber nonlinear architecture, providing a compact, scalable, and deployable platform for future fiber-based nonlinear and quantum photonic technologies.

**Funding**

This work was supported by the National Research Foundation (RS-2024-00438839; 2022M3H4A1A04096396) and the Institute for Information & Communications Technology Planning & Evaluation (IITP) Grant (RS-2025-25464832) of Korea.

**Acknowledgements**

The authors thank the UNIST Central Research Facilities (UCRF) QuantumNanoFab supported by NRF (RS-2024-00401037) and IITP ITRC center (RS-2023-00259676).

**Data Availability Statement**

The data that support the findings of this study are available from the corresponding author upon reasonable request.

Supporting Information

**Toward Compact Fiber In-line Nonlinear Devices via Highly Efficient Nanophotonic Cavity Interface**

*Mahsa Haddadi Moghaddam[1,†], Kirlie Iulius Figuera Michal[1,†], Sangwoo Lee[1], Sijin Sung[1], Jongwon Lee[2], Hyeong-Ryeol Park[1], and Je-Hyung Kim[1,3]**

[1]Department of Physics, Ulsan National Institute of Science and Technology, Ulsan 44919, Republic of Korea

[2]Department of Electrical Engineering, Ulsan National Institute of Science and Technology, Ulsan 44919, Republic of Korea

[3]Graduate School of Quantum Science and Technology, Ulsan National Institute of Science and Technology, Ulsan 44919, Republic of Korea

.

**S1 - The simulated structural parameters of the cavity**

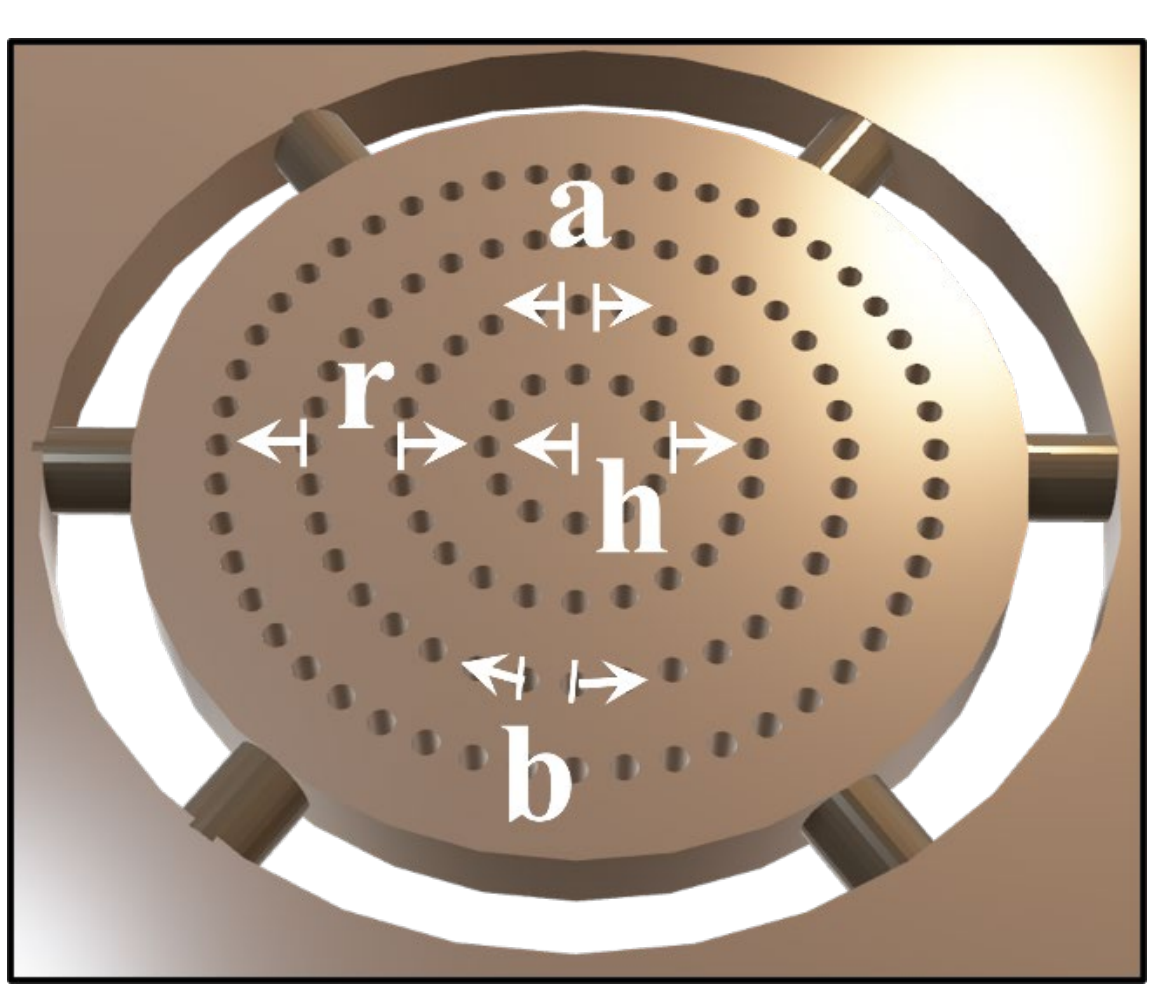

**Figure S1.** Schematics of the h-CBG. The simulated geometries are defined by several key structural parameters, including the radial period between adjacent holes (r), the radius of the central disk (h), the hole size (a), and the axial period between holes (b).

## S2 - Sample Fabrication

The devices were fabricated on commercially available GaN-on-silicon wafers consisting of a 350 nm GaN layer on a 200 nm AlN buffer. A 100 nm Cr hard mask was first deposited and patterned by electron-beam lithography, followed by a $Cl_2/O_2$ dry etch to define the mask features. The GaN cavities were subsequently etched using a $Cl_2$/Ar plasma process. After fabrication, the GaN/AlN structures remained too strongly bonded to the silicon substrate to allow direct transfer onto the optical fiber. To enable transfer, a selective wet-etching process was carried out to remove the underlying substrate beneath the fabricated devices. Specifically, the silicon substrate and AlN buffer were selectively etched using HNA and hot phosphoric acid, respectively, forming air-suspended membranes (area (I)). This isotropic etching process releases the GaN patterns from the silicon substrate along the device edges, while the larger surrounding regions remain attached to the silicon substrate (area (III)). During this step, sprue-like support structures prevent device loss by maintaining mechanical connection to the parent flake. With continued selective wet etching, the undercut extends further around the h-CBG patterns, creating a freestanding GaN region surrounding the structures, as shown in area (II). [1,2]

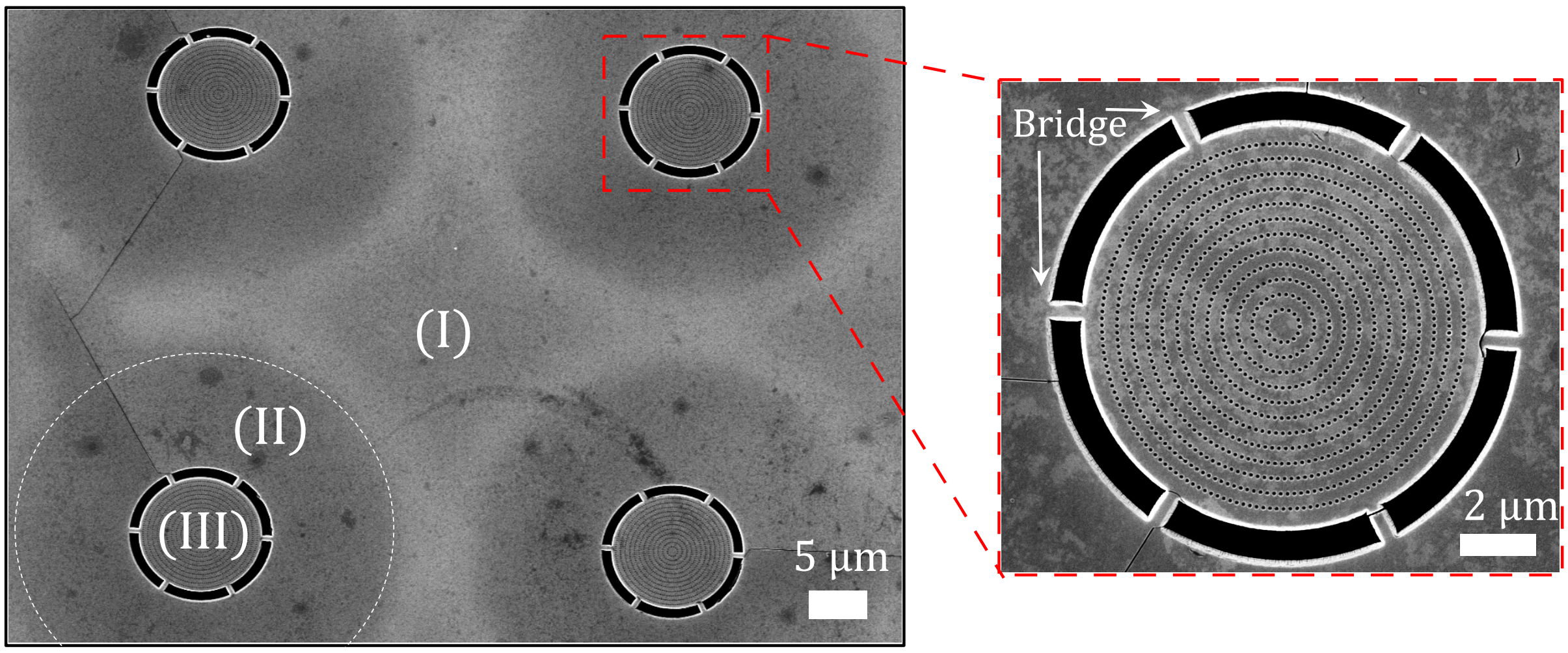

**Figure S2.** SEM image of the fabricated sample showing three distinct regions: (I) GaN on the silicon substrate, (II) the free-standing GaN membrane, and (III) the GaN h-CBG cavity. The inset shows a magnified SEM image of a single fabricated h-CBG cavity.

## S3 - Cross-polarized reflectivity setup for cavity-mode measurement

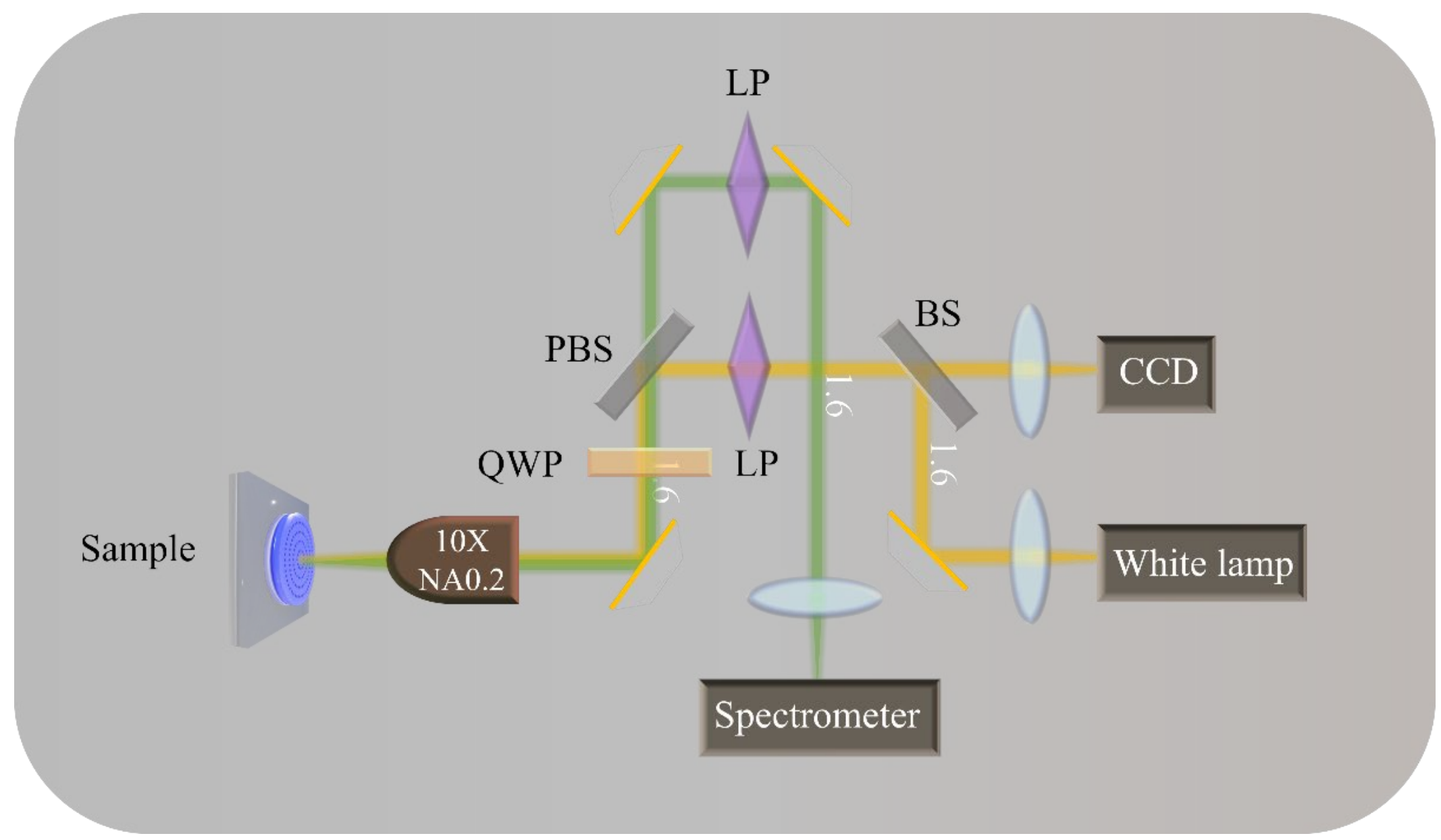

**Figure S3.** Cavity resonances were measured using a cross-polarized reflectivity scheme. Broadband illumination from a white lamp was coupled into the optical path through a beam splitter and linearly polarized before being directed toward the sample through a Polarized beam splitter (PBS), linear polarizer, and a quarter-wave plate (QWP). The beam was focused onto the cavity by a 10× objective (NA = 0.2). Upon reflection from the sample, the light traversed the QWP again and the linear polarizer, resulting in a polarization rotation relative to the incident beam. As a consequence, the reflected cavity signal was redirected by the PBS into the detection arm, while the co-polarized background was largely rejected. The output was coupled into the fiber core using a lens. The light emerging from the fiber was then collimated with a 40× objective and directed either into a spectrometer (SpectraPro HRS-750 Princeton Instruments) to measure the reflected spectra, or to a cooled Si charge-coupled device (CCD) camera to monitor the sample position and reflected image.

**S4- The optimized cavity geometries**

| GaN h-CBG | Radial period $r$ (nm) | Central disk radius $R$ (nm) | Hole radius $a$ (nm) | Axial period $l$ (nm) | Mode $\lambda$ (nm) |
|---|---|---|---|---|---|
| **A** | **373** | **483.4** | **59.7** | **208.9** | **778.06** |
| **B** | **372** | **482.1** | **63.2** | **212.0** | **784.53** |
| **C** | **371** | **480.8** | **61.2** | **211.5** | **786.02** |
| **D** | **372** | **482.1** | **59.5** | **212.0** | **788.98** |
| **E** | **370** | **482.9** | **59.9** | **185.0** | **790.97** |
| **F** | **372** | **485.5** | **55.8** | **212.0** | **793.98** |

**Table S4.** Geometric parameters and resonance wavelengths of the GaN h-CBG cavities. The optimized cavity design targeting resonance near 784 nm consists of a central disk radius $R$ of approximately 482 nm, a radial spacing between nanohole arrays $r$ of approximately 372 nm, an axial spacing between adjacent nanoholes $l$ of approximately 212 nm, and a nanohole radius $a$ of approximately 63 nm, with the membrane thickness fixed at 350 nm.

## S5 – SHG reflection measurement setup

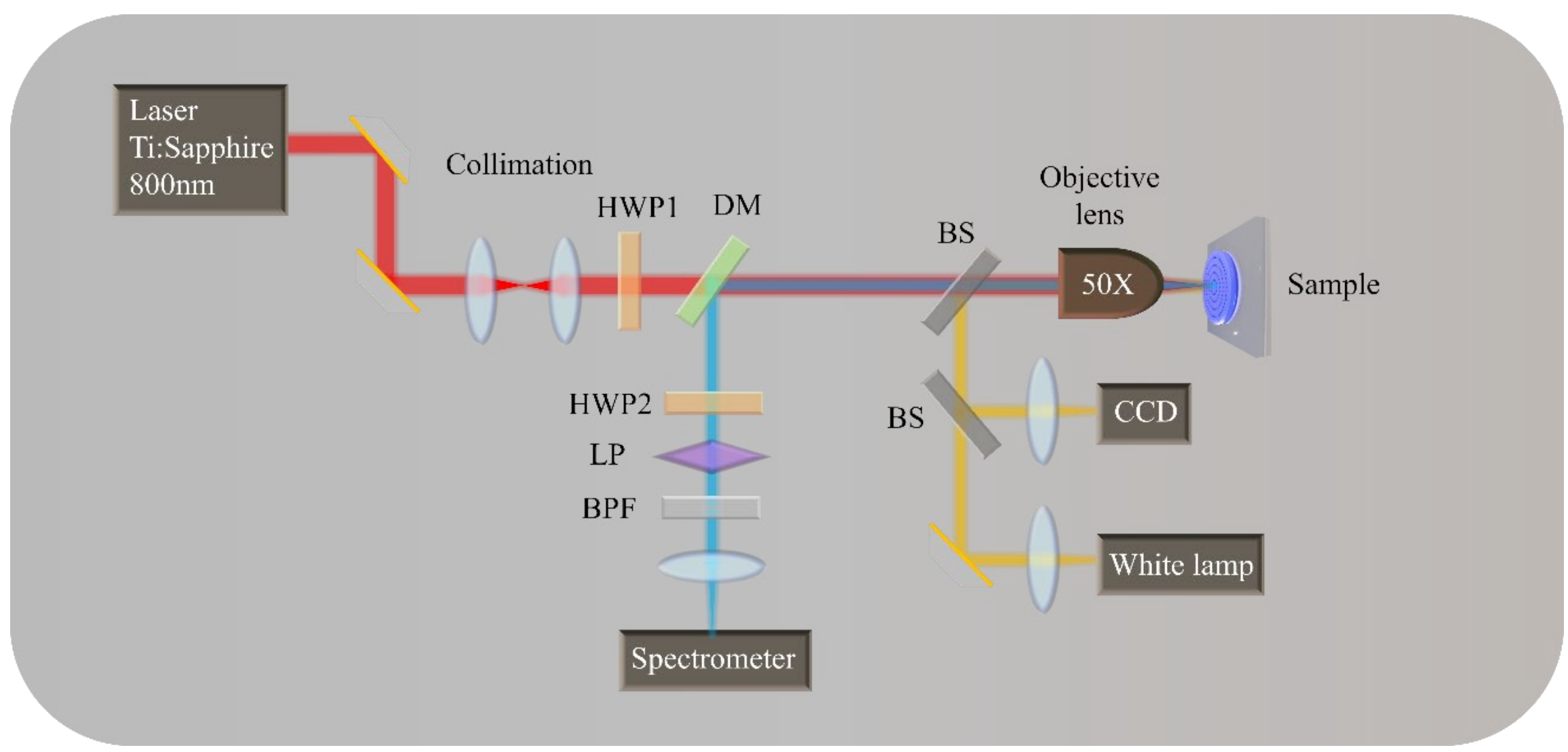

**Figure S5.** The second-harmonic generation signal was characterized using a reflection-type optical setup. A Ti:Sapphire laser operating at 785 nm with a repetition rate of 80 MHz and a pulse duration of 130 fs was used as the fundamental excitation source. The pump beam was first collimated and passed through a half-wave plate (HWP1) to control the incident polarization before being directed to the sample and focused using a 50× objective lens. The generated SHG signal, centered near 395 nm, was separated from the reflected fundamental beam by the dichroic mirror (DM) and directed into the detection arm. In this path, a second half-wave plate (HWP2) and a linear polarizer (LP) were used to analyze the polarization of the SHG emission, while a band-pass filter (BPF) was employed to suppress residual pump light and isolate the SHG signal. The filtered signal was then focused on a spectrometer for spectral analysis. To facilitate sample positioning and alignment, a white lamp illumination path and a CCD imaging arm were incorporated into the same system through beam splitters (BS). This allowed simultaneous visual monitoring of the sample while performing SHG measurements. Using this reflection configuration, the SHG spectra from the cavity structures could be measured efficiently while maintaining accurate focusing and alignment on the target device.

**S6 - Conversion efficiency calculation**

The SHG conversion efficiency was estimated from the measured average pump power and the detected average SHG output power. In these measurements, the excitation pump is a pulsed laser with an average pump power of 250 mW, a pulse duration of 130 fs, and a repetition rate of 80 MHz. Because SHG is a nonlinear process that depends on the instantaneous optical intensity rather than only on the average power, it is important to convert the average pump power into the corresponding pulse energy, peak power, and peak intensity at the focal spot.

The absolute SHG conversion efficiency was calculated as $\eta_{abs} = P_{2\omega}/P_{\omega}$, and the power-normalized efficiency as $\eta_p = P_{2\omega}/P_{\omega}^2$ . In this experiment, the pump peak intensity is estimated from the average pump power $P_\omega$=93 mW , repetition rate f=80 MHz, pulse width $\tau$=130 fs, and focal spot diameter around 8.12 μm, using $E_p = P_\omega/f$ , $P_{peak} = E_p/\tau$ , and $I_p = P_{peak}/[\pi(\frac{d}{2})^2]$. [3-6]

For h-CBG-B, with $P_{2\omega}$=31.3 μW, the estimated peak intensity is 17.3 $GWcm^{-2}$, the absolute conversion efficiency is $3.37\times10^{-4}$, the power-normalized efficiency is $3.62\times10^{-3}$ $W^{-1}$ (0.362% $W^{-1}$), and the intensity-normalized efficiency is $2.09\times10^{-4}$ $cm^2GW^{-1}$.

For the fiber-integrated h-CBG-B, under the same pump conditions and with a measured SHG power of $P_{2\omega}$=4.6 μW. The corresponding absolute conversion efficiency is $4.95\times10^{-5}$, the power-normalized efficiency is $5.32\times10^{-4}$ $W^{-1}$, and the intensity-normalized efficiency is $3.08\times10^{-5}$ $cm^2GW^{-1}$.

**S7 – Stability of the SHG signal**

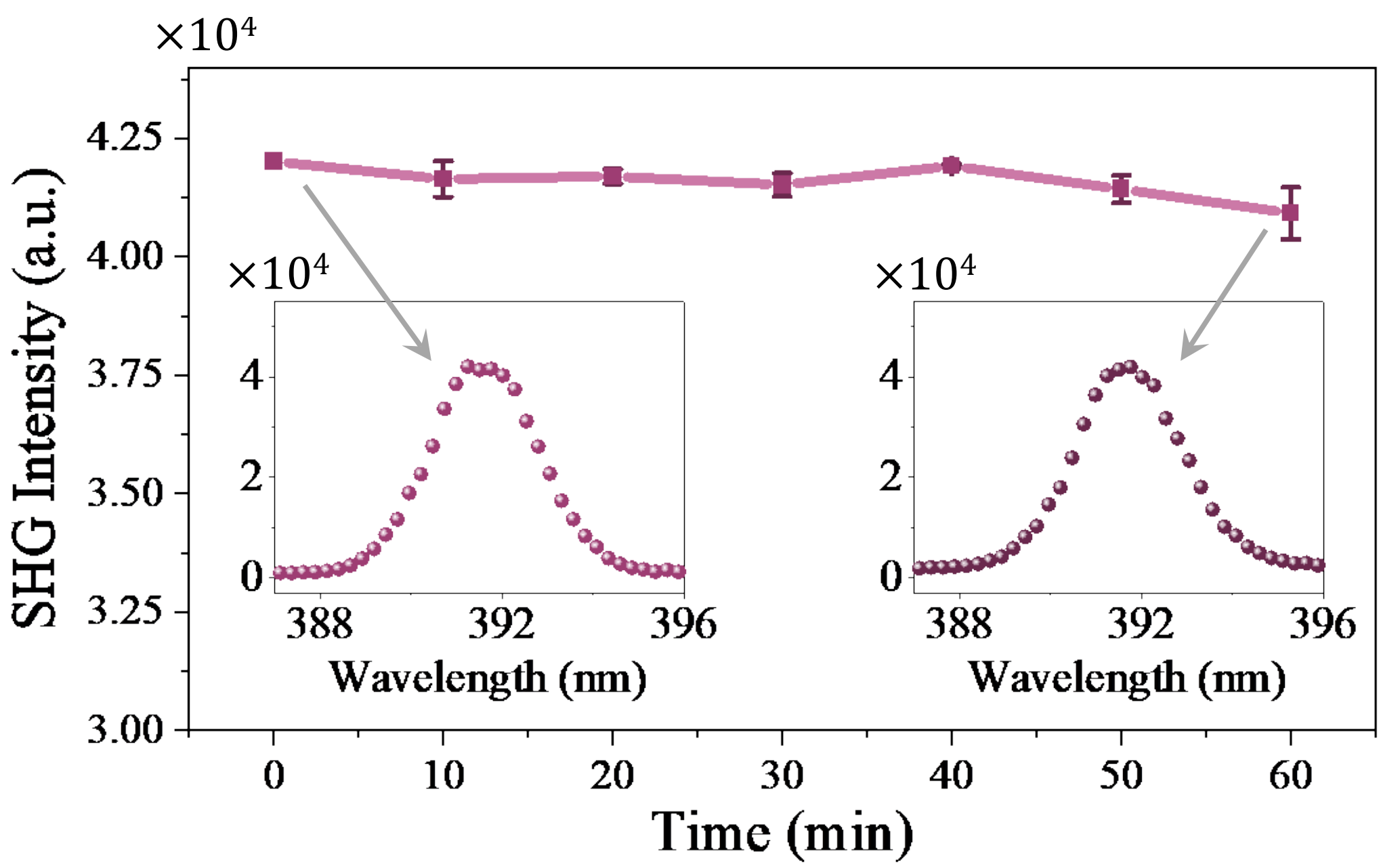

**Figure S7.** High-stability SHG under continuous irradiation (Pp=93 mW). Error bars show the SHG intensity fluctuations over 1 hour compared to the average intensity, and the inset shows the SHG intensity evolution.

## S8 - Deterministic transfer of the h-CBG onto optical fibers

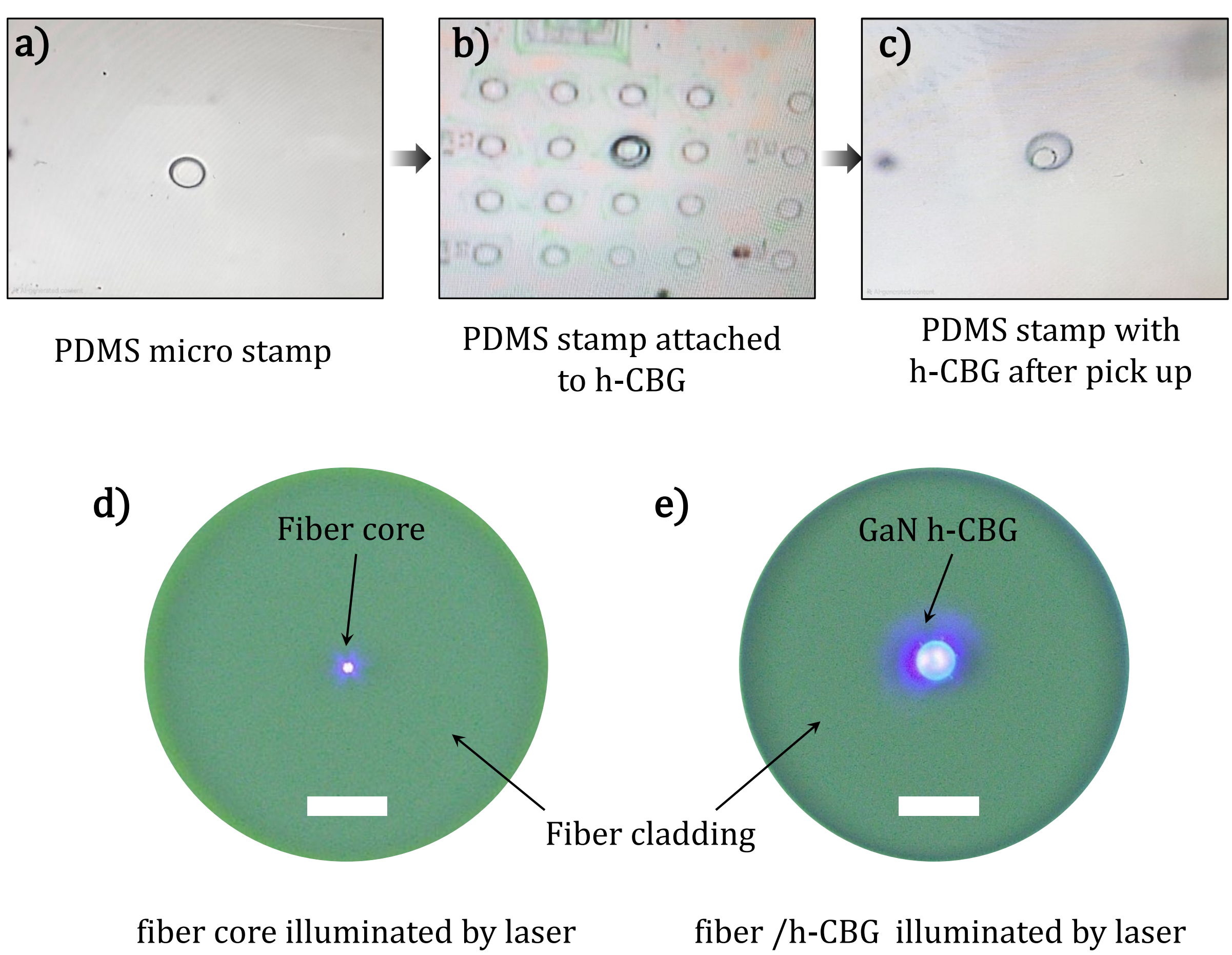

**Figure S8**. (a) Optical image of the PDMS microstamp. (b,c) Optical microscope images of the fiber-integration process, (b) showing alignment of the PDMS microstamp with the h-CBG cavity and **(c)** pickup of the h-CBG device for subsequent transfer onto the fiber facet. Optical microscope images of the fiber core illuminated by a 403 nm laser coupled from the opposite end of the fiber **(d)** before and **(e)** after transferring the h-CBG cavity onto the fiber core. (All scale bars in the optical microscope images of the fiber facets are 20 μm.)

To demonstrate a fiber-integrated h-CBGh-CBG device, we employed a deterministic pick-and-place transfer technique using a polydimethylsiloxane (PDMS) micro stamp. The PDMS stamps were fabricated from a silicon mold prepared by photolithographically patterning an array of 20 μm diameter circles, followed by deep reactive-ion etching to transfer the pattern into silicon to a depth of 20 μm. PDMS was then prepared from a Sylgard 184 elastomer kit with a base-to-curing-agent mass ratio of 10:1. The mixture was spin-coated onto the silicon mold and cured at 100 °C for 1 h, resulting in transparent PDMS microstamps mounted on glass slides (Figure

S8a shows an optical image of the PDMS stamp). The entire transfer process was monitored in real time under an optical microscope. As shown in Figure 3a and 3d, the fabricated h-CBG devices were released as air-suspended membranes supported by thin bridge structures connected to the parent wafer. Owing to its compliant and adhesive surface, the transparent PDMS micro stamp could selectively contact the suspended h-CBG region (Figure S8b), detach it from the substrate (Figure S8c), and transfer it with high precision onto the core of the optical fiber. To align the cavity with the fiber core, a blue laser beam at 403 nm was launched from the opposite end of the fiber, providing a visual guide for centering the device on the fiber facet (Figure S8d). The h-CBG device remained firmly attached to the fiber facet through van der Waals adhesion and was accurately centered on the fiber core (Figure S8e). In these experiments, we used a single-mode patch fiber (Thorlabs **SM300**, pure silica core, wavelength range 320~430 nm, FC/PC connector, and 1 m length) with a mode field diameter of 2 μm and numerical aperture (NA) of 0.12 and ultra-high NA fiber (Thorlabs **UHNA3**, pure silica core, wavelength range 900~1600 nm, FC/PC connector, and 30 cm length). For comparison and further measurements, a single-mode patch fiber (Thorlabs S630-HP, pure silica core, wavelength range 630~860 nm, FC/PC connector, and 1 m length) with a mode field diameter of 4 μm and numerical aperture (NA) of 0.12 was also used.